# Growth of Ba$_2$CoWO$_6$ Single Crystals and their Magnetic, Thermodynamic and Electronic Properties

Abanoub Hanna[1,2], A.T.M.N. Islam[2], C. Ritter[3], S. Luther[4], R. Feyerherm[2], B. Lake[1,2]

[1]Institut für Festkörperphysik, Technische Universität Berlin, Germany

[2]Helmholtz-Zentrum Berlin für Materialien und Energie GmbH, Berlin, Germany

[3]Institut Laue-Langevin, Grenoble Cedex 9, France

[4] Hochfeld-Magnetlabor Dresden (HLD-EMFL), Helmholtz-Zentrum Dresden-Rossendorf, Dresden, Germany

**Abstract**

This study explores the bulk crystal growth, structural characterization, and physical property measurements of the cubic double perovskite Ba$_2$CoWO$_6$ (BCWO). In BCWO, Co$^{2+}$ ions form a face-centered cubic (FCC) lattice with non-distorted cobalt octahedra. The compound exhibits long-range antiferromagnetic order below $T_N$ = 14 K. Magnetization data indicated a slight anisotropy along with a spin-flop transition at 10 kOe , a saturation field of 310 kOe and an ordered moment of 2.17 µB at $T$ = 1.6 K,. Heat capacity measurements indicate an effective $j$ = 1/2 ground state configuration, resulting from the combined effects of the crystal electric field and spin-orbit interaction. Surface photovoltage analysis reveals two optical gaps in the UV-Visible region, suggesting potential applications in photocatalysis and photovoltaics. The magnetic and optical properties highlight the significant role of orbital contributions within BCWO, indicating various other potential applications.

**Introduction**

Double perovskite oxides (A$_2$BB'O$_6$) have distinct structural properties and a high degree of order, which initiated a wide range of fundamental research, such as in the fields of quantum magnetism and colossal magnetoresistance [1-6]. The various possible combinations of paramagnetic cations (B, B') in double perovskites offer a playground for magnetism. The arrangement of magnetic ions in the B or B' sites form a face-centred cubic (FCC) sublattice of the magnetic ions coordinated octahedrally by oxygen (BO$_6$).

The FCC lattice is well known to host various magnetic states, such as antiferromagnets of type-I (MnTe$_2$ [7]), type-II (MnO[8]), type-III (MnS$_2$ [7]), and spiral (helical) (γ-Fe [9]) magnetic structures. The wealth of these magnetic structures is partly explained by the inherent geometric frustration of the FCC lattice [10, 11]. Theoretical calculations suggested that the FCC lattice can host other interesting magnetic quantum states due to the competition between the nearest three possible exchange interactions, $J_1$, $J_2$ and $J_3$ [12, 13].

Ba$_2$CoWO$_6$ (BCWO) is a cubic double perovskite where Co$^{2+}$ ions form an FCC lattice (Fig. 1). Single crystals of BCWO were previously reported synthesized by the flux method using BaCl$_2$ flux resulting in crystals of maximum size ≈ 10×7×7 mm$^3$ [14, 15]. BCWO crystallizes in the space group *Fm-3m* (No. 225) with high symmetry and ideal octahedral coordination of the cobalt ions [16, 17].

Magnetic measurements of BCWO indicated long-range antiferromagnetic order below the Néel temperature $T_N$ = 18 K [16, 17]. The effective moment $\mu_{eff}$ = 5.18 $\mu_B$ was considerably larger than the spin-only value of 3.87 $\mu_B$ for $S = 3/2$, implying an unquenched orbital contribution. This orbital contribution may lead to anisotropic magnetic behaviour in BCWO, as demonstrated by high-field magnetization data [14]. The Curie-Weiss temperature $\Theta_{cw}$ = - 75 K (fitted range 100-330 K) and in other reports, $\Theta_{cw}$ = - 41 K (fitted range 200-300 K) indicated that BCWO might indicate magnetic frustration [16-18]. The frustration index $\Theta_{cw}/T_N$ is 4.2 and 2.3 respectively, implying possible suppression of magnetic order due to competing interactions. The magnetic structure was determined using neutron diffraction which indicated that the spins propagate along the magnetic vector $k = (1/2,1/2,1/2)$ with an ordered moment of 2 $\mu_B$ at 2 K [17]. The spins were found to order ferromagnetically in the triangular planes perpendicular to the [111] direction, with antiferromagnetic ordering between adjacent planes [17, 19].

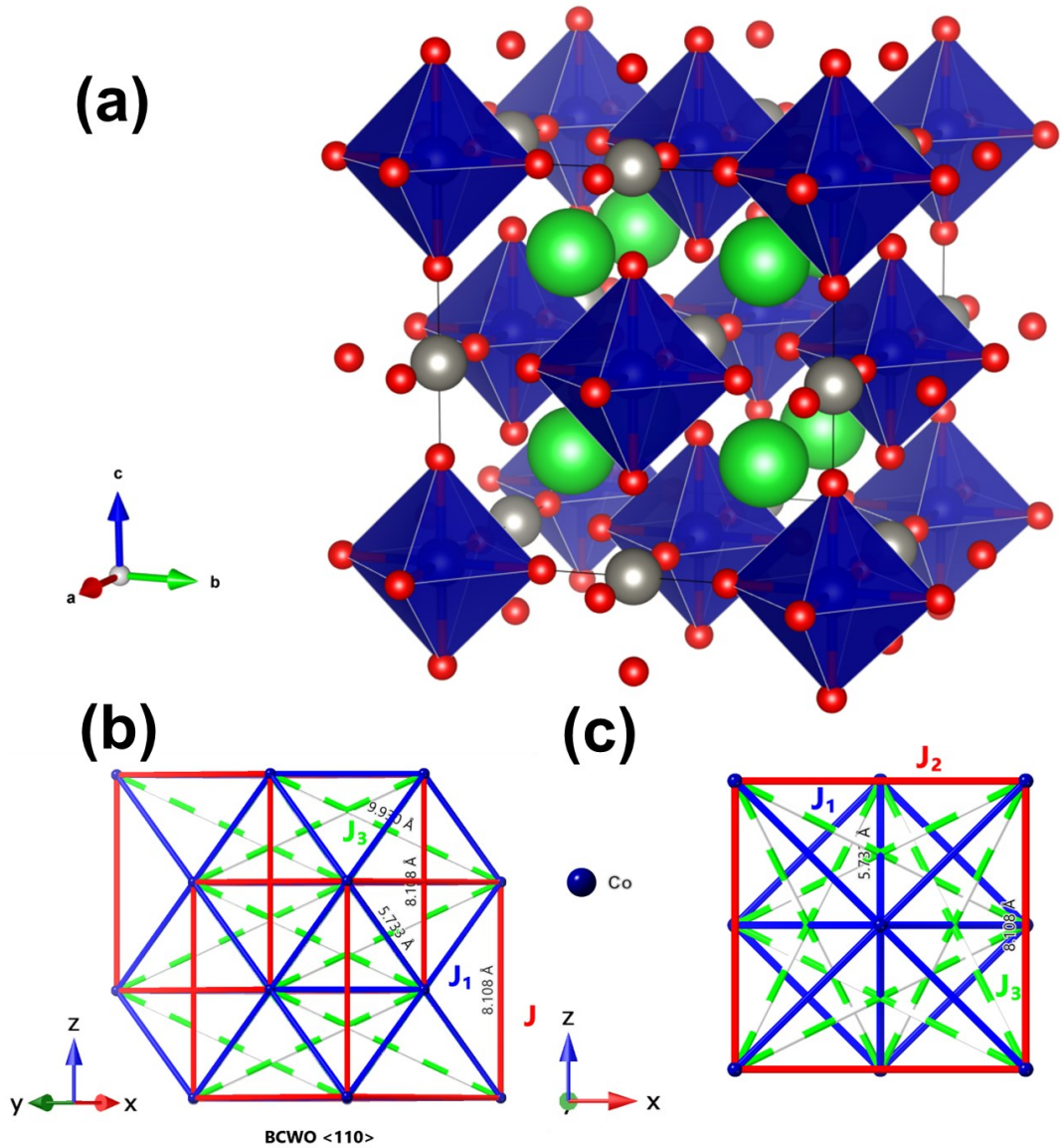

Fig. 1 The Crystal structure of $Ba_2CoWO_6$ (BCWO) consists of $Co^{2+}$ ions (blue) surrounded by six $O^{2-}$ ions (red), Ba (green) and W (grey) ions. (b,c) The arrangement of cobalt ions in the FCC lattice forms edge-shared tetrahedra. Possible exchange interactions between first, second and third neighbours are denoted by $J_1$(blue), $J_2$ (red) and $J_3$ (green).

Here, we introduce the crystal growth and characterization of the physical properties of a bulk single crystal of Ba$_2$CoWO$_6$ grown by the floating zone method. Physical characterization reveals a small anisotropy attributed to Cobalt ions. The slight anisotropy in the magnetization data is accompanied by a spin-flop transition at 1 T. High-field magnetization data indicate a saturation field of 31 T and a saturation moment of 2.1 $\mu_B$ at 2 K. Heat capacity measurements elucidate that the ground state has an effective $j = 1/2$ configuration, as commonly observed for Co$^{2+}$ ions in octahedral environment, indicating the significant role of the combined crystal electric field and spin-orbit coupling in this system.

**Experimental**

To grow single crystals of BCWO, a stoichiometric powder was synthesized by solid-state reaction as previously described [16, 19-21]. The precursors (BaCO$_3$, CoO, WO$_3$) were mixed in a stoichiometric ratio and heated in several steps to 900°C, 1200°C and 1350°C in the air for 12 hours at each temperature followed by intermediate grindings. BCWO was obtained as a dark brown powder.

A single crystal of BCWO was grown using the optical floating zone (FZ) technique. The crystal growth was performed using an image furnace (Crystal Systems Corp., FZ-T 10000-H-VI-VPO) at the CoreLab Quantum Materials at Helmholtz-Zentrum Berlin für Materialien und Energie (HZB). High-density feed rods were prepared by packing BCWO powder into rubber tubes and subjecting them to a pressure of 2000 bars in a cold isostatic press (CIP) machine, yielding a uniform cylindrical rod with dimensions of 70 mm length and 5 mm diameter. The rod was sintered in the air for 12 hours at 1350°C and used as a feed for crystal growth. The feed rod was introduced to the image furnace with 4 lamps of 300 W and a small part of the rod was used as a seed to initiate the growth.

BCWO is a congruently melting compound. Therefore, the growth experiment was performed in air at ambient conditions with 78% of the lamp power, and the growth rate was 8 mm/h. The growth experiment lasted around 8 h and the grown crystal shows clear shiny facets on the surface and has dimensions of 55 mm length and 6 mm diameter as shown in Fig. 2(a).

The structural characterization and phase purity were conducted using both X-ray powder diffraction (PXRD) by Bruker D8 XRD diffractometer equipped with copper source Kα X-ray (wavelength λ =1.542 Å) at HZB, and neutron powder diffraction (NPD) at D2B neutron diffraction spectrometer with incident neutrons wavelength λ =1.594 Å at Institut Laue-Langevin (ILL), France. Both measurements were performed on crushed single crystals at room temperature. For NPD measurement, the sample weight was 3.5 g.

Furthermore, the single crystal was systematically investigated and mapped using a Laue Diffraction Camera, which was also used to align the crystallographic directions for physical properties measurements. The Laue spots were solid and clear, representing the high quality of the single crystal as shown in Fig. 2 (b, c).

The physical properties of the single crystal of BCWO were analysed using both the Physical Property Measurement System (PPMS) and SQUID magnetometer (MPMS) at HZB. The magnetic susceptibility, magnetization and heat capacity were measured up to 90 kOe over a temperature range of 2 - 400 K. High field magnetization was measured on single crystals for different directions at $T$ = 1.6 - 20 K using a compensated pickup-coil system in a pulse field magnetometer in a 4He flow-cryostat for fields up to 60 T at Dresden High Magnetic Field Laboratory, Helmholtz-Zentrum Dresden-Rossendorf (HLD-HZDR).

The optical properties of the compound were probed using surface photovoltage (SPV) spectroscopy at room temperature [22]. SPV was measured by the fixed capacitor method using a chopper with a frequency of 8 Hz for both single crystal and powder in the photon energy range of 0.5 – 5 eV.

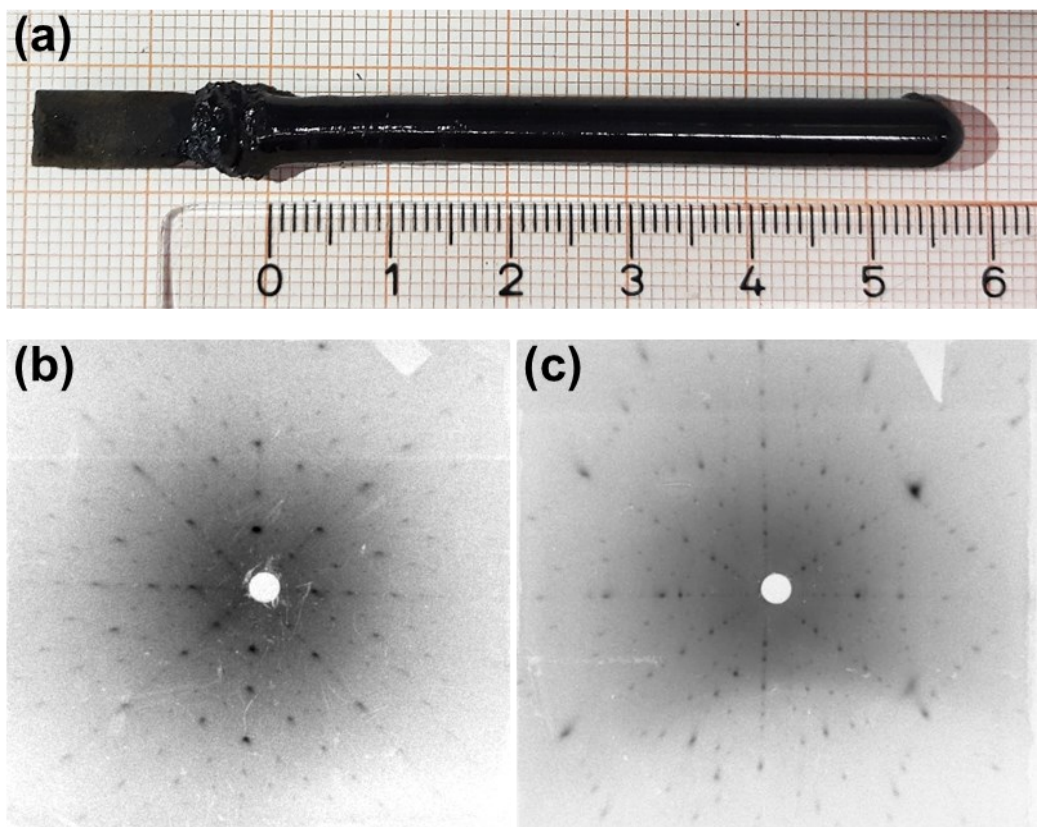

Fig. 2 (a) Optical photograph of BCWO single crystal (57 mm length and 5 mm diameter) grown in air at ambient pressure. (b, c) X-ray Laue backscattering photographs of the grown crystal nearly along the [100], and [110] directions, respectively.

## Results

### 1 Structural characterization

The structural parameters and phase purity of the poly and single-crystalline phases were first inspected by room-temperature powder X-ray diffraction (PXRD). For more precise quantitative analysis, the single crystal was also analysed using neutron powder diffraction (NPD). Both PXRD and NPD data were refined by the *FullProf* software using Rietveld refinement to extract the structural parameters and the volume fraction of each phase [23].

The PXRD data of the polycrystalline sample revealed the presence of extra peaks attributed to the secondary phase $BaWO_4$ as shown in Fig 3. The main phase of $Ba_2CoWO_6$ was refined by the *Fm-3m* space group (No. 225). The minor impurity phase $BaWO_4$ was refined by the $I4_1/a$ space group (No. 88). The refinement of the PXRD of the powder and crystal revealed that the amount of $BaWO_4$ was reduced from 1.5% in the powder to 0.7 % in the crystal.

Table 1 Refinement parameters for BCWO in the *Fm-3m* space group (No.225) at room temperature including the lattice parameter (*a*) and the profile R-factor (*Rp*). The Wyckoff sites of the elements are Ba(*8c*), Co(*4b*), W (*4a*) and O(*24e*) are fixed by symmetry except for the atomic position of oxygen in the *x* direction ($O_x$) which is also listed.

|  | $BaWO_4$ (%) | *a* (Å) | *Rp* | $Chi^2$ | $O_x$ |
| --- | --- | --- | --- | --- | --- |
| **Powder-XRD** | 1.5 | 8.108(1) | 5.44 | 2.01 | 0.234(5) |
| **Crystal-XRD** | 0.7 | 8.108(1) | 2.41 | 3.13 | 0.246(3) |
| **Crystal-NPD** | 0.9 | 8.106(1) | 3.59 | 2.28 | 0.237(1) |
| **Powder-NPD** [17] | 5 | 8.108(1) | 4.54 | 1.87 | 0.2625(1) |

The presence of the impurity $BaWO_4$ implies the presence of unreacted quantities of CoO and BaO. These impurities are not detectable by conventional PXRD due to their low concentrations and the similarity in their crystallographic properties to BCWO. Their overlapping peaks with BCWO in the PXRD spectra prevent clear identification and differentiation. It should be noted that the amount of $BaWO_4$ is smaller than the previously reported powder data [17]. This impurity phase was also used in the refinement of the NPD pattern, although it did not improve the refinement and its amount is less than 0.9%, revealing the improved purity of the single crystal compared to the initial powder.

The refinement parameters are summarized in Table 1 for the powder, single crystal, and a previous report. The refinement confirmed the cubic symmetry of BCWO with a rock-salt order consisting of regular $BaO_{12}$ cuboctahedra and corner-sharing regular $CoO_6$ and $WO_6$ octahedra. $Ba^{2+}$ occupies the *8c* Wyckoff position, while $Co^{2+}$ occupies the *4b* position, $W^{6+}$ occupies the *4a* position and $O^{2-}$ occupies the *24e* position. These atomic positions are fixed by symmetry except for oxygen. The occupancy of Ba, Co and W does not change significantly from ideal stoichiometry if allowed to vary, and the NPD data which is sensitive to Oxygen did not find any deficiency. The NPD data also implied the absence of site mixing of $Co^{2+}$ and $W^{6+}$ ions. These results reveal the high quality of the BCWO crystal and the undistorted octahedral geometry of the $Co^{2+}$ ions.

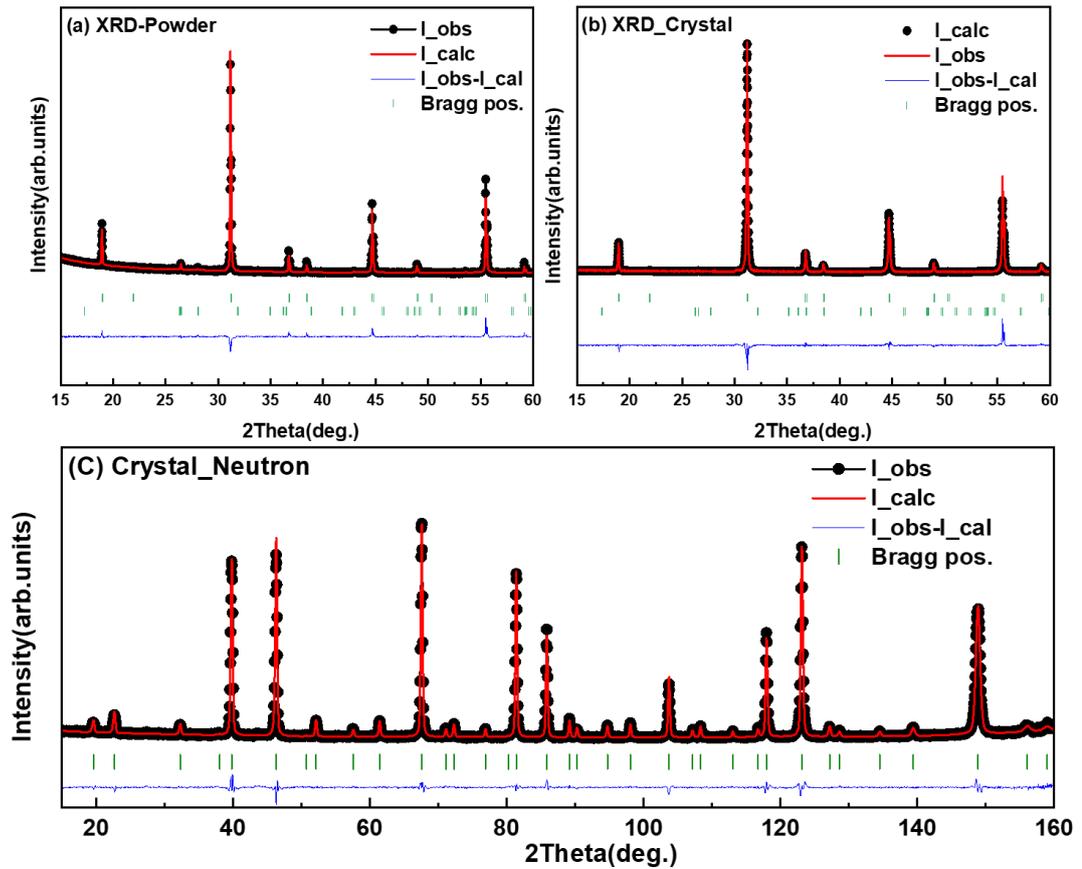

Fig. 3 (a, b) Rietveld refinement of the PXRD of BCWO (a) powder and (b) single crystal. The refinement involves both the BCWO and BaWO$_4$ phases. (c) Rietveld refinement of NPD of BCWO grinded crystal. The observed and calculated intensities are represented by the solid black circles and red lines respectively, and the difference between them is given by the blue solid line. The Bragg peak positions for the *Fm-3m* space group of BCWO are given by the first set of vertical green lines while the second set of green lines is for the *I4$_1$/a* space group of BaWO$_4$.

## 2 Magnetization

Fig. 4 represents the temperature dependence of the magnetic susceptibility of the BCWO crystal for applied magnetic fields along the [110], [100] and [111] directions. The magnetic susceptibility exhibits a broad maximum of around 18 K for all field directions. The antiferromagnetic ordering temperature, $T_N$ = 14.0(2) K for $H$ = 10 kOe, was determined by the maximum of the first derivative of the susceptibility data as shown in Fig. 4(b). The susceptibility data exhibits slight direction dependence, revealing a weak anisotropy below $T_N$ where the susceptibility is lower for [110] compared to [111] and [100]. Anisotropic behaviour is also observed for the lower applied field of 1kOe as shown in Fig. 4(c) where below $T_N$ the susceptibility value is now lower for [100] and [111] compared to [110], in contrast to the results at 10 kOe (Fig. 4(a))

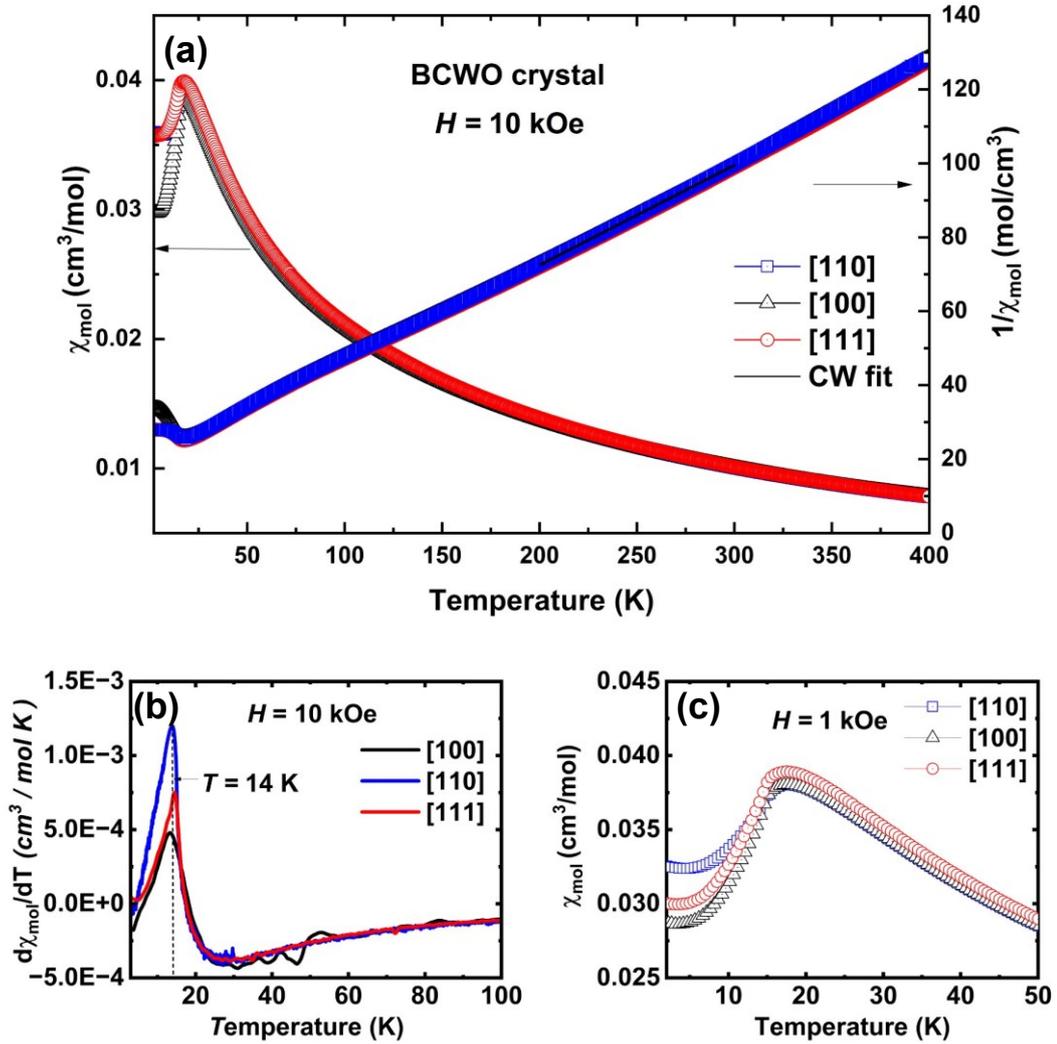

Fig. 4 (a) Temperature dependence of magnetic susceptibility (left axis) and inverse susceptibility (right axis) of BCWO for different crystallographic directions at $H$ = 10 kOe over the temperature range 2 – 400 K. (b) First derivative of magnetic susceptibility at 10 kOe indicating the Néel temperature (c) Magnetic susceptibility of BCWO for 1 kOe applied magnetic field along the [100], [110] and [111] directions over the temperature range 2 - 50 K. The field dependence of the magnetic susceptibility reveals the anisotropy for different directions at 1kOe and 10 kOe below $T_N$.

For fitting the inverse susceptibility data according to the modified Curie-Weiss (CW) model, we restricted the temperature range to 200 - 300 K since the slight curvature in the inverse susceptibility data in the range between 30-150 K may be attributed to short-range magnetic order or a change in the population of excited states of the $Co^{2+}$ ions [17]. Also, if the $T$-range is above 300 K, a partial population of higher excited states may affect the temperature dependence of the susceptibility. The temperature-independent diamagnetic susceptibility was calculated from the diamagnetic correction of each element as $\chi_0$ = -1.50 x$10^{-4}$ $cm^3$/mol and was included as a constant in the model [24]. The fitting results are summarized in Table 2.

Table 2 Fitted values for the inverse susceptibility of the BCWO crystal according to the Curie-Weiss model for the temperature range 200-300 K for various field strengths applied along the crystallographic directions [100], [110] and [111]. Where $H$ is the applied magnetic field, $\Theta_{CW}$ is the Curie Weiss temperature, and $\mu_{eff}$ is the effective moment.

| $Ba_2CoWO_6$ 200-300 K | $H = 1$ kOe | | $H = 10$ kOe | |
|---|---|---|---|---|
| | $\Theta_{CW}$ [K] | $\mu_{eff}$ [$\mu_B$] | $\Theta_{CW}$ [K] | $\mu_{eff}$ [$\mu_B$] |
| [100] | -73.7(2) | 5.45(1) | -70.8(2) | 5.45(2) |
| [110] | -68.9(2) | 5.40(3) | -66.2(2) | 5.45(3) |
| [111] | -75.5(3) | 5.50(3) | -68.1(1) | 5.44(3) |

The CW fitting of the single crystal data gives $\Theta_{CW}$ between -69.5 K and -77.5 K, and effective moment $\mu_{eff}$ = 5.40 - 5.50 $\mu_B$ for the temperature range 200 - 300 K, as summarized in Table 2. These values are in good agreement with previous reports [12,13]. The effective moment is significantly larger than the spin-only moment of 3.87 $\mu_B$, suggesting that the orbital contribution to the total magnetic moment is partially unquenched. Such a behaviour is commonly observed for $Co^{2+}$ ions in an octahedral environment.

The lowest-energy electron structure, governed by Hund's rules, of a free $Co^{2+}$ ion is $^4F$ ($S = 3/2$, $L = 3$). Under a regular octahedral crystal electric field (CEF) this state splits and the ground state, denoted as $^4T_{1g}$, is 12-fold degenerate (orbital triplet × spin quartet). In this state, the orbital moment is quenched and the spin-only effective moment is $\mu_{eff}$ = 3.87 $\mu_B$. The effect of spin-orbit coupling (SOC) further splits the degenerate energy levels and partially restores the orbital momentum. The final state is a set of Kramers doublets ($\Gamma_6$ and $\Gamma_7$) and two $\Gamma_8$ quartets, which consequently are associated with effective spin $j = 1/2$ and $j = 3/2$, respectively. For $Co^{2+}$ in a perfect octahedral environment, the lowest energy level is the $\Gamma_6$ doublet, i.e., a $j = 1/2$ ground state is realized. The $\Gamma_6$ state has the g factor $g = 4.33$, with an effective moment of about 3.8 $\mu_B$ [25]. The effective moment rises with temperature due to the population of excited CEF+SOC states, resulting in a crossover to values $\mu_{eff} > 5$ $\mu_B$ above 150-200 K.

Isothermal magnetization measurements provide further evidence for the weak anisotropy of BCWO. Fig. 5 (a, b) denotes the magnetization curves and the first field derivative of the magnetization respectively for different directions. The measurements show that the magnetization of BCWO has non-linear behaviour for the [100] and [111] directions and the first derivative shows a peak at 10 kOe representing a spin-flop transition. The spin flop transition indicates a weak anisotropy in the system which favours certain spin ordering directions below $T_N$. When a magnetic field is applied that is strong enough to overcome the anisotropy, the AFM-ordered magnetic spins rotate to a direction perpendicular to the field direction to minimize their total energy [26].

The spin-flop transitions observed for [100] and [111] imply that the spins have components along these directions. For the [110] direction, the change in the derivative of magnetization is much more gradual and shows no sign of a spin-flop transition suggesting that the spins aligned perpendicular to it. Therefore, the spins point between the [111] and [100] axis while being perpendicular to the [110] axis. These findings agree with the previous reports stating that the magnetic spins are tilted by 23° ±5° from the [111] axis, as proposed by Cox *et.al.* [19].

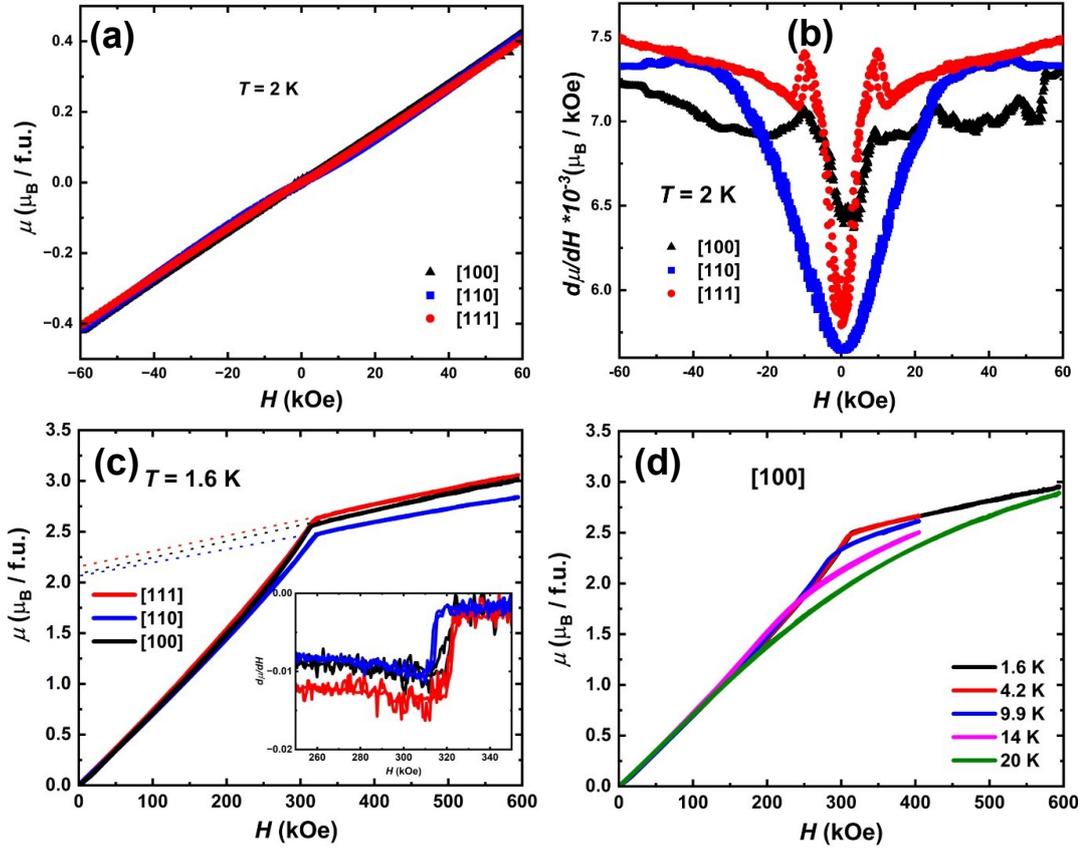

Fig. 5. Isothermal magnetization of BCWO: (a) Magnetic moment per formula unit (f.u.) as a function of the magnetic field applied along the [111], [100] and [110] directions of the BCWO crystal. (b) The field derivative of the field-induced moment per f.u. highlights the spin-flop around 10 kOe for the [100] and [111] directions. (c) High field magnetization of different directions of BCWO crystal at $T$ = 1.6 K revealing saturation at the field $\mu_0 H$ = 310 - 320 kOe. The dashed lines represent the Van-Vleck paramagnetism extrapolated to zero fields. After subtracting this contribution, the saturation moment is about 2.1 $\mu_B$. (d) Variation of magnetization with temperature for the [100] direction elucidating the nonlinear magnetization and absence of saturation above $T_N$.

Fig. 5 (c) shows the high field magnetization measurements for fields up to $H$ = 600 kOe along the [111], [100] and [110] directions of BCWO at $T$ = 1.6 K. The saturation field is indicated by the change of slope of magnetization curves at 310 - 320 kOe as shown in the inset. The linear increase of magnetization above the saturation field is due to the Van-Vleck paramagnetic behaviour of $Co^{2+}$ ions. After subtraction of the Van-Vleck contribution, saturation moments between 2.05 $\mu_B$ (for [100] and [110]) and 2.13 $\mu_B$ (for [111]) are observed. These values are in good agreement with the ordered magnetic moment of 2 $\mu_B$ at 2 K from neutron diffraction measurements [17] and with the value $g$ = 2.17 $\mu_B$ expected from an effective spin-1/2 model with $j =1/2$ and $g = 4.33$ for $Co^{2+}$ with perfect octahedra [27, 28]. Above the Néel temperature, the magnetization is nonlinear and does not reach saturation up to 600 kOe as shown in Fig. 5 (d).

**3 Heat Capacity**

The main evidence for the $j = 1/2$ ground state in BCWO originates from the heat capacity measurement and entropy calculations. The heat capacity of BCWO is shown in Fig. 6(a) in zero magnetic field and for field applied along the [110] direction for the temperature range $T$ = 2 - 120 K. The measurement reveals a second-order phase transition at 14 K attributed to the long-range antiferromagnetic order in agreement with the magnetic susceptibility data. No significant change was noticed in the heat capacity under different applied magnetic fields up to 90 kOe and the Néel temperature is almost field-independent. We modelled the phonon contribution to the heat capacity for the temperature range 50 – 180 K with the sum of one Debye mode and one Einstein mode (with equal weight) according to the equation

$$c_p(T) = 3Rn \left( \frac{1}{2}\left(\frac{T}{\theta_D}\right)^3 3 \int_0^{\frac{\theta_D}{T}} \frac{x^4 e^x}{(e^x - 1)^2} dx + \frac{1}{2}\left(\frac{\theta_E}{T}\right)^2 \frac{e^{\theta_E/T}}{(e^{\theta_E/T} - 1)^2} \right)$$

Where $R$ is the universal gas constant, $n$ is the number of ions per formula unit which is $n = 10$ for BCWO, and $\theta_D, \theta_E$ are the Debye and Einstein temperatures, respectively. The model suggests a Debye temperature $\theta_D = 264(1)$ K and Einstein temperature $\theta_E = 539(2)$ K. The phonon contribution was extrapolated to low temperatures and subtracted from the measured heat capacity to give the magnetic contribution. The extracted magnetic heat capacity $c_{mag}$ and entropy changes are shown in Fig. 6(b).

The magnetic entropy was determined by integrating $c_{mag}/T$ over temperature up to 100 K. It increases in the range from 2 K to ~50 K, where it reaches a value close to $R\ln2$ per $Co^{2+}$, where $R\ln2$ is the expected value for the $j = 1/2$ ground state configuration. Therefore, this result confirms the effective $j = 1/2$ Kramers doublet of the $Co^{2+}$ ions in the ground state of BCWO. The entropy gained up to the magnetic phase transition is only 0.6 $R\ln2$ indicating that the spins are strongly correlated above the Néel temperature, which is a feature of frustrated magnetic systems. Along with the magnetic susceptibility and magnetization data, BCWO can be considered a weakly frustrated magnet with a slight anisotropy of the cobalt ions.

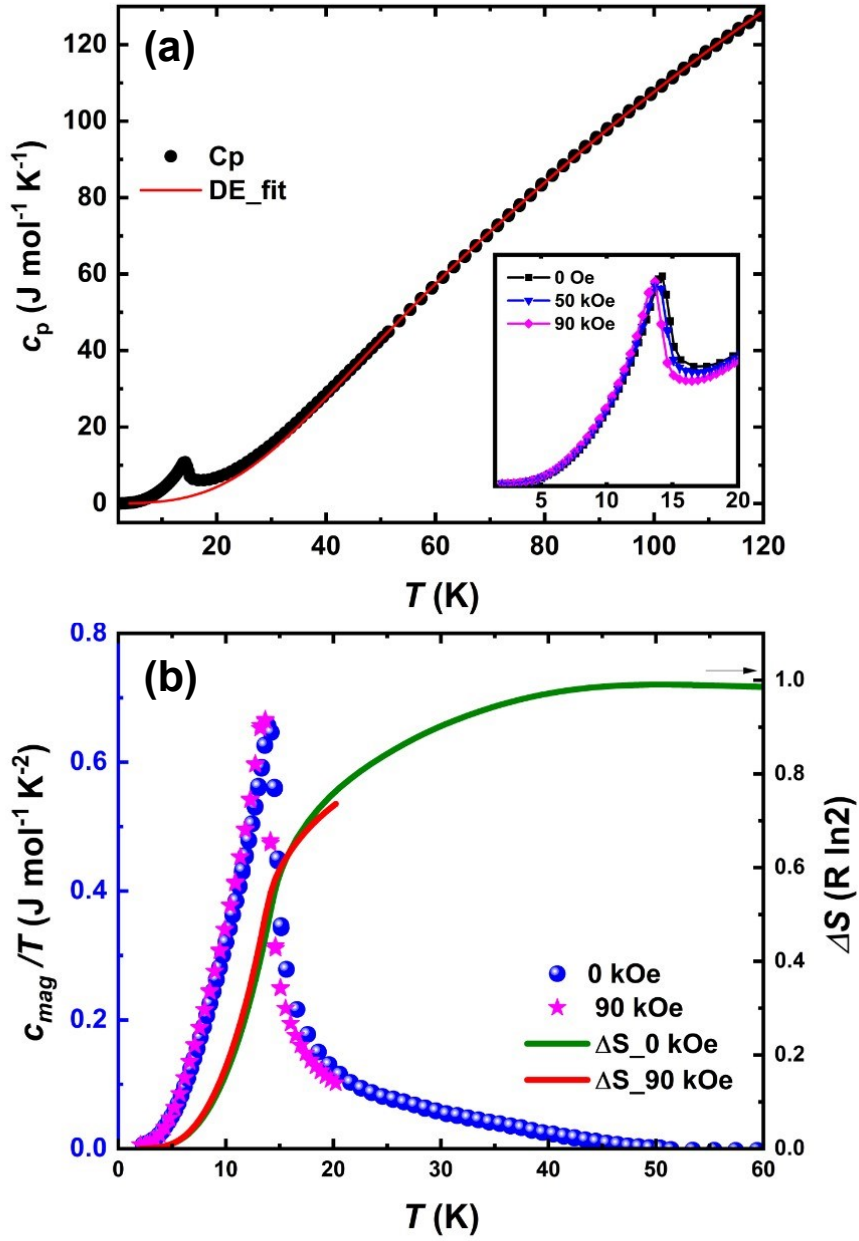

Fig. 6 (a) Heat capacity of BCWO crystal in zero fields (black circles) and the phononic contribution of the heat capacity (red line) modelled by the Debye- Einstein terms. Inset represents the magnetic field dependence of the heat capacity up to 90 kOe along the [110] direction. (b) The extracted magnetic contribution of the heat capacity over temperature for zero field and 90 kOe. The change in entropy is shown on the left axis in units of $R\ln2$, confirming the $j = 1/2$ ground state.

# 4 Electronic structure

Several theoretical reports calculated the electronic and magnetic band structures of BCWO [16, 20, 29]. These reports suggested that BCWO is half-metallic due to the electron-electron interactions between the $Co^{2+}$ ions. The reported UV-visible spectroscopy measurements indicated an optical bandgap of BCWO of approximately 2.46 eV [20]. To compare our results with previously measured data, we measured the surface photovoltage (SPV) of both the polycrystalline and single-crystal samples at room temperature for the energy range 0.5 - 5 eV as shown in Fig. 7. Our measurements showed two gaps in both the single crystal at 2.60(5) eV and 3.44(5) eV (21000, 27700 cm$^{-1}$) and in the powder pellet at 2.32(5) eV and 3.23 (5) eV. These values are depicted from the deviation from the linear part of the energy dependence of the spectra as shown in Fig.7 (a). These two electronic states might represent the possible excited states resulting from the $d$-orbital splitting due to the crystal field. These energy values suggest the possible transitions between the ground state $^4F\,^4T_{1g}$ and excited states $^4F\,^4A_{2g}$ and $^4P\,^4A_{2g}$. The crystal field splitting in BCWO is close to the value which is expected as a result of octahedral ligand field splitting of the $^4F$ and $^4P$ terms of the single-ion Co(II) (19000,22500 cm$^{-1}$) [30]. In this case, the second excited state may represent a transition from the high-spin state (HS) $t_{1g}(5)\,e_g(2)$ to the low-spin state (LS) $t_{1g}(6)\,e_g(1)$, which would involve an electron transfer to another orbital. The spectrum also revealed consistency with the previously reported data for the value of the optical gap around 2.5 eV. This optical gap supports the theoretical calculations which suggested that the ground state is half-metallic [20, 21, 29]. SPV shows rather semiconducting properties from the value of the bandgaps and supports the high spin electronic configuration of $Co^{2+}$ ($3d^7$) and $W^{6+}$($5d^0$)[29].

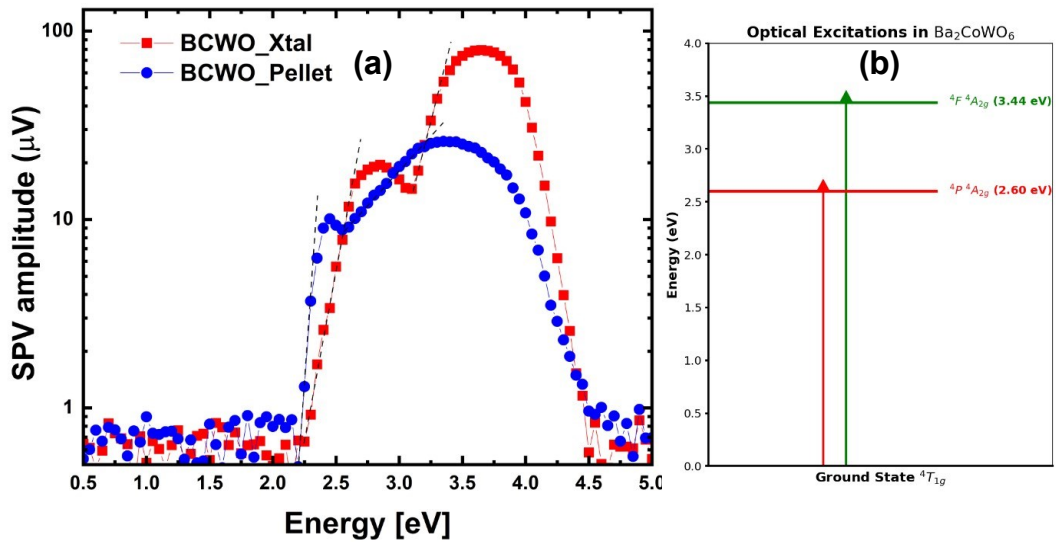

Fig. 7 (a) Surface photovoltage measurement on the surface of BCWO crystal and pressed pellet showing that BCWO has two optical band gaps. For the crystal: 2.60 (5) and 3.44 (5) eV. For the pellet: 2.32 (5) and 3.23 (5) eV. The dashed black line indicates the deviation from the linear part of the absorption. (b) Schematic description of the corresponding excited states of BCWO.

## Conclusions

In this work, we introduced the bulk crystal growth and physical properties of the double perovskite FCC lattice antiferromagnet $Ba_2CoWO_6$. Our investigation revealed that BCWO has a magnetic ground state with a $j = \frac{1}{2}$ configuration. The presence of unquenched orbital momentum due to strong spin-orbit coupling in the system is found along with weak frustration depicted from the magnetic susceptibility data. The magnetization data revealed the existence of a small anisotropy and a spin flop transition at 10 kOe due to the collinear antiferromagnetic structure, as well as a lower saturation moment than expected for $Co^{2+}$ due to the combination of CEF+SOC effects. The orbital contribution seems to be rather significant affects the magnetism, and has the potential to induce complex magnetic features in the system. The surface photovoltage measurements confirmed that BCWO has two optical gaps of 2.6, and 3.44 eV and possesses a configuration of $Co^{2+}W^{6+}$. The two gaps in the UV-visible range suggest many interesting applications of BCWO as an optically active material for photocatalysis and photovoltaic applications.

## Conflict of interest

The authors declare no conflict of interest.

## Acknowledgements

We acknowledge the CoreLab Quantum Materials at Helmholtz Zentrum Berlin für Materialien und Energie (HZB), Germany, where the powder and single crystal samples of $Ba_2CoWO_6$ were synthesized and measured. The authors acknowledge the support of Hochfeld Magnetlabor Dresden at Helmholtz Zentrum Dresden Rossendorf (HLD-HZDR), a member of the European Magnetic Field Laboratory (EMFL). This work is also partly based on experiments performed at the reactor neutron source Institut Laue-Langevin, Grenoble, France. We acknowledge Sahana Rößler and Miguel Carvalho from the Max-Planck-Institute for Chemical Physics of Solids, Dresden for sharing their X-ray absorption spectroscopy results. We acknowledge Igal Levine and Thomas Dittrich from the Solar Energy Division, HZB for performing the surface photovoltage measurements.